\documentclass{JHEP}
\usepackage{epsfig}

\newcommand{\be}{\begin{equation}}
\newcommand{\ee}{\end{equation}}
\newcommand{\bea}{\begin{eqnarray}}
\newcommand{\eea}{\end{eqnarray}}
\newcommand{\bean}{\begin{eqnarray*}}
\newcommand{\eean}{\end{eqnarray*}}

\newtheorem{proposition}{\sf PROPOSITION}

\def\ket#1{\left| #1\right\rangle}
\def\bra#1{\left\langle #1 \right|}  

 \newcommand{\IR}{{\bf R}}

\preprint{MIT-CTP-3113\\  {\tt hep-th/0105024}}

\title{The Identity String Field and the Tachyon Vacuum}

\author{Ian Ellwood, Bo Feng, Yang-Hui He and Nicolas Moeller
\footnote{
Research supported in part by
the CTP and LNS of MIT and the U.S. Department of Energy 
under cooperative research agreement \# DE-FC02-94ER40818.
I. E. is supported in part by an NSF Graduate Fellowship.
N. M. and Y.-H. H. are also supported by the Presidential Fellowship
of MIT.}
\\
Center for Theoretical Physics,
\\Massachusetts Institute of Technology,\\ Cambridge, MA 02139, USA\\
\email{iellwood,fengb,yhe,moeller@ctp.mit.edu}
}

\abstract{We show that
the triviality of the entire cohomology of the new BRST operator
$Q$ around the tachyon vacuum is
equivalent to the $Q$-exactness of the identity ${\cal I}$ of the
$\star$-algebra. We use level truncation to show that as the level is
increased, the identity becomes more accurately $Q$-exact. We carry
our computations up to level nine, where an accuracy of $3\%$ is
attained. Our work supports, under a new light, 
Sen's conjecture concerning the absence of open string degrees of
freedom around the tachyon vacuum. As a by-product, a new and simple
expression for ${\cal I}$ in terms of Virasoro operators is found.
}
\begin{document}
\section{Introduction}
Regarding the fate of the tachyon in various systems such as
brane-antibrane pairs in Type II theories as well
as the D25-brane in the bosonic string theory, Sen proposed 
his famous three conjectures in \cite{9902105, 9904207}. 
These state that
(i) The difference in energy between the perturbative
	and the tachyon vacuum exactly cancels
	the tension of the corresponding D-brane system;
(ii) After the tachyon condenses, all open string 
	degrees of freedom disappear, leaving us with the closed string 
	vacuum; and 
(iii) Non-trivial field configurations 
	correspond to lower-dimensional D-branes.

Because tachyon condensation is an off-shell process\footnote{For
	some early works concerning tachyon condensation please
	consult \cite{Halpern}.
	}, we must use the
formalism of string field theory. 
Both Witten's cubic open string field theory 
\cite{WITTENSFT} and his background independent open string
field theory \cite{9208027,9210065,9303143,9311177} 
seem to be good candidates.
Indeed, in the last two years, there has been a host of works aimed 
to understand Sen's three conjectures by using the above two string field
theories as well as the non-linear sigma-model (Born-Infeld action)
\cite{0011033}.
Thus far, Sen's 
first and third conjectures have been shown to be true to 
a very high level of
accuracy (\cite{KS} - \cite{0008127}); 
they have also been proven
analytically in Boundary String Field Theory
(\cite{0009103} - \cite{0009191}). 
The second conjecture however, is still puzzling.

Let us clarify the meaning of this conjecture. From a physical point
of view, after the 
tachyon condenses to the vacuum, the corresponding D-brane system
disappears and there is no place for open strings to end on. Therefore
at least all perturbative conventional open
string excitations (of ghost number 1) should decouple from the
theory. There has been a lot of work to check this statement,
for example (\cite{0008033}-\cite{ShatashviliTalk}). In particular,
using level truncation, \cite{0103085} verifies that
the scalar excitations at even levels (the $Q$ closed scalar fields)
are also $Q$-exact to very high accuracy.

However, as proposed in \cite{0012251,0102112} there is a little
stronger version for the second conjecture. There,
Rastelli, Sen and Zwiebach suggest that after a field redefinition, 
the new BRST operator may be taken\footnote{
The first String 
Field Theory action with pure ghost kinetic operator
was written down in \cite{NewExact}.}
to be simply
$c_0$, or more generally a linear combination of operators
of the form $(c_n + (-)^n c_{-n})$. For such a new BRST operator,
not only should the
conventional excitations of ghost number 1 disappear, 
but more precisely the full cohomology of any ghost number
of the new BRST operator around the tachyon vacuum 
vanishes\footnote{An evidence
	for the triviality of a subset of the discrete ghost number one 
	cohomology was presented recently in \cite{0103103} which
	complemented \cite{0103085}.}.
Hence these authors propose that Sen's second conjecture should hold
in such a stronger level. In fact, Sen's second conjecture 
suggests also that around the tachyon vacuum, there should be
only closed string dynamics. However, we will not touch upon
the issue of closed strings
in our paper and leave the reader to the references
\cite{0103056,9206084,closed,0012081}.

Considering the standing of the second conjecture, it is the aim of
this paper to address to what degree does it hold, i.e., 
whether the cohomology of $Q_{\Psi_0}$ is trivial only for ghost number 1
fields  or for fields of any ghost 
number. We will give evidence which shows that
the second conjecture holds in the strong sense, and is 
hence consistent with the proposal in \cite{0012251,0102112}.

Our discussion relies heavily upon the existence of a string
field ${\cal I}$ of ghost number 0 which is the identity of the 
$\star$-product. It satisfies 
\[
{\cal I} \star \psi = \psi \star {\cal I} = \psi
\]
for any state\footnote{There are some mysteries regarding of the
	identity. For example, in \cite{0006240} the authors showed that 
	this identity string field is 
	subject to anomalies, with consequences that $\cal{I}$ may be 
	the identity of the $\star$-algebra only on a subspace of the whole 
	Hilbert space. In the following, we will first assume that $\cal{I}$ 
	behaves well on the whole Hilbert space, and postpone some
	discussions thereupon to Section 4.
	}
$\psi$. The state ${\cal I}$ was first 
constructed in the oscillator basis in \cite{gross-jevicki, 
gj2}. Then a recent work \cite{0006240} gave a recursive way of constructing 
the identity in the (background independent) total-Virasoro basis
which shows its universal property in string field theory.
As a by-product of our analysis, 
we have found a new and elegant analytic expression for
${\cal I}$ without recourse to the complicated recursions.

Ignoring anomalies, the fact that $Q_{\Psi_0}$ is a derivation of the
$\star$-algebra implies that ${\cal I}$ is 
$Q_{\Psi_0}$ closed and the problem is to determine
whether it is also $Q_{\Psi_0}$ exact, i.e., if there exists
a ghost number $-1$ field $A$, such that ${\cal I}=Q_{\Psi_0} A$.
If so, then for an arbitrary $Q_{\Psi_0}$ closed state $\phi$
we would have
\[
\begin{array}{rcl}
Q_{\Psi_0} (A \star \phi) 
& = &(Q_{\Psi_0} A) \star \phi - A \star (Q_{\Psi_0} \phi) \\
& = &\cal{I} \star \phi \\
& = & \phi,
\end{array}
\]
where in the second step, we used the fact that $\phi$ is 
$Q_{\Psi_0}$-closed, 
and in the last step, that ${\cal I}$ acts as the identity on $\phi$. 
This means that any $Q_{\Psi_0}$-closed field $\phi$ is also
$Q_{\Psi_0}$-exact, in other words,
the entire cohomology of $Q_{\Psi_0}$ is trivial. 

Therefore we have translated the problem of the triviality of
the cohomology of $Q_{\Psi_0}$ into the issue of the exactness of the
identity ${\cal I}$. 
In this paper, we will use the level truncation method
to show that the state $A$ indeed exists for the tachyon vacuum $\Psi_0$
up to an accuracy of $3.2 \%$.

The paper is structured as follows. In Section 2, we explain 
the above idea of the exactness of ${\cal I}$ in detail. 
In Section 3, we use two different methods to find
the state $A$: one without gauge fixing and the other, in the
Feynman-Siegel
gauge. They give the results up to an accuracy of $2.4 \%$ and $3.2 \%$
respectively. In Section 4, we discuss the behaviour of ${\cal I}$
under level truncation and perform a few consistency checks on our
approximations.
Finally, in Section 5 we make some concluding
remarks and address some further problems and directions.

A few words on nomenclature before we proceed.
By $\ket{0}$ we mean the $SL(2,\IR)$-invariant vacuum and
$\ket{\Omega} := c_1 \ket{0}$. We consider $\ket{\Omega}$ to be
level 0 and hence $\ket{0}$ is level 1. Furthermore, in this paper 
we expand our
fields in the universal basis (matter Virasoro and ghost oscillator modes). 
\section{The Proposal}
To reflect the trivial cohomology of the BRST operator at the
stable vacuum, Rastelli, Sen and
Zwiebach \cite{0102112} proposed that after a field redefinition,
the new BRST operator $Q_{new}$ may be taken to be simply $c_0$, 
or more generally
a linear combination of operators of the form $(c_n+(-)^n c_{-n})$.
For such operators, 
there is an important fact: there is an operator $A$ such that
\[
\{A, Q_{new} \}=I,
\]
where $I$ is the identity operator.
For example, if
$Q_{new}=c_n+(-)^n c_{-n}$, we can choose $A=\frac12(b_{-n}+(-)^n b_{n})$
because 
$
\{ \frac12(b_{-n}+(-)^n b_{n}), Q_{new} \} =
\{ \frac12(b_{-n}+(-)^n b_{n}), c_n+(-)^n c_{-n}\} =1.
$
Therefore, if the state $\Phi$ is closed, i.e., $Q_{new} \Phi=0$, then
we have
\begin{equation}
\begin{array}{ccl}
\Phi & = & \{ A,Q_{new}\} \Phi
        = AQ_{new}\Phi+ Q_{new}A\Phi \\
     & = & Q_{new}(A\Phi)
\end{array}
\end{equation}
which means that $\Phi$ is also exact.  Thus the existence of such an
$A$ guarantees that the cohomology of $Q_{new}$ is trivial. 

In fact the converse is true.  Given a $Q_{new}$ which has vanishing cohomology
we can always construct an $A$ such that $\{A, Q_{new} \} = I$.  Suppose
that we denote the string Hilbert space at ghost level $g$ by $V_g$.  
Define the subspace $V_g^C$ as the set of all closed elements
of $V_g$.  We can then pick a complement, $V_g^N$, to this 
subspace\footnote{More precisely, the space $V_g$ could be split as
	$V_g=V_g^C \oplus (V_g/V_g^C)$ where $(V_g/V_g^C)$ is a
	vector space of equivalence classes under the addition of
	exact states. $V_g^N$ should be considered as a space of
	representative elements in $(V_g/V_g^C)$.}
satisfying $V_g = V_g^C \oplus V_g^N$. Note that it consist
of vectors which are not killed by $Q_{new}$.
This subspace $V_g^N$,  
is not gauge invariant but any specific choice
will do. The important point is that because we have assumed that $Q_{new}$ has
no cohomology, the restriction of $Q_{new}$ to $V_g^N$ given by
\[
  Q_{new} \Bigr|_{V_g^N} :V_g^N \to V_{g+1}^C,
\]
has no kernel and is surjective\footnote{As remarked in the previous
	footnote, if we use $(V_g/V_g^C)$ instead of $V_g^N$, the
	mapping is an isomorphism of vector spaces.}
on $V_{g+1}^C$.  Thus it has an inverse which
we denote $A$
\[
A \Bigr|_{V_{g+1}^{C}} 
    \equiv Q_{new}^{-1} \Bigr|_{V_{g+1}^{C}} : V_{g+1}^{C} \to V_g^N.
\]
This insures that on the space $V_{g}^{C}$, $\{ A ,Q_{new} \}=I$ holds since
if $\Phi$ is $Q_{new}$-closed, 
\[
  \{ A ,Q_{new} \} \Phi = A Q_{new} \Phi + Q_{new} A \Phi 
      = Q_{new} Q_{new}^{-1} \Phi = \Phi.
\]

The above discussion only defines the action of $A$ on $V_g^C$, what
remains is to define its action on the complement $V_{g}^{N}$. 
Here there is quite a bit of freedom since one can choose any map that takes 
$V_{g}^{N}$ into $V_{g-1}^{C}$.  Assuming this, we have that for  
$\Phi \in V_{g}^{N}$, 
\[
  \{ A ,Q_{new} \} \Phi = A Q_{new} \Phi + Q_{new} A \Phi
      = Q_{new}^{-1} Q_{new} \Phi + Q_{new}^2 \chi = \Phi,
\]
where by assumption $A \Phi$ is $Q_{new}$-closed (because it is
in $V_{g-1}^{C}$) and thus equals $Q_{new} \chi$
for some $\chi \in V_{g-2}^{N}$.  In general one can insist that $A$ satisfies
more properties.  For example if we set $A |_{V_{g}^{N}} = 0$ we get that 
$A^2=0$. We summarize the above discussion as
\begin{proposition}
The cohomology of $Q_{new}$ is trivial iff there exists an operator
$A$ such that $\{A ,Q_{new} \}=I$.
\end{proposition}

The basic hypothesis of this paper is that not only does such an operator $A$
exist for $Q_{\Psi_0}$, 
but also for special choices of $A$, {\em the action of $A$ can be expressed
as the left multiplication by the ghost number $-1$ string field} which we
denote as $A \star$.  
Thus we are now
interested in satisfying the equation $\{A\star ,Q_{new}\} = I$.  
Writing this
out explicitly we have
\bean 
  \{A\star ,Q_{new}\} \Phi &=& A \star (Q_{new}\Phi) + Q_{new}(A\star\Phi) \\
  &=& A \star Q_{new}(\Phi) + (Q_{new}A) \star \Phi - A \star (Q_{new} \Phi)\\
  &=& (Q_{new}A) \star \Phi.
\eean
In order for the last line to equal $\Phi$ for all $\Phi$ we need that
\begin{equation}\label{E:QAeqI}
  Q_{new} A = \mathcal{I},
\end{equation}
where $\mathcal{I}$ is the identity of the $\star$-algebra.

For the case of interest, we wish to study the physics around the minimum
of the tachyon potential.  We recall that for a state $\Phi$,
the new BRST operator around the solution $\psi$ of the EOM is given by
\begin{equation}
Q_{\psi} \Phi = Q_B (\Phi)+ \psi\star (\Phi)-(-)^{\Phi} (\Phi)\star \psi.
\end{equation}
Using this expression for the BRST operator we can rewrite the 
basic equation (\ref{E:QAeqI}) as
$Q_{\psi} A = Q_B (A)+ \psi\star (A)+ (A)\star \psi = \mathcal{I}$.
For general vacua $\psi$, such a string field $A$ will not exist. For
example in the perturbative vacuum, $\psi=0$, $Q_{\psi}$ is simply
$Q_B$. It is easy to show here that there is no solution for $A$ because
the $Q_B$ action preserves levels while
${\cal I}$ has a component at level one (namely $\ket{0}$), 
but the minimum level of a
ghost number $-1$ state $A$ is 3. Indeed, for a more general solution
$\psi\neq 0$ (such as the tachyon vacuum), the star product will not
preserve the level and so it may be possible
to find $A$. Our endeavor will be to use the level truncation scheme
to find $A$ for the tachyon vacuum $\Psi_0$, i.e., to find a solution
$A$ to the equation 
\begin{equation} \label{condition}
Q_{\Psi_0} A = \mathcal{I}.
\end{equation}
%There is another issue we need to mention. Recalling the definition
%of the operator $A$, we see that there is much freedom therein.
%For example, different choices of $V_g^N$ (recall that
%$V_g^N$ is a representative element) will give different actions of
%$A$ on $V_{g+1}^C$. Even if we fix the choice of $V_g^N$, we still have
%some freedom when we map $V_g^N$ to $V_{g-1}^C$. For our special choice
%of $A$ in (\ref{E:QAeqI}), it is easy to see that the freedom can be
%written as 
Note that this equation is invariant under
\[
A \rightarrow A + Q_{\Psi_0} B
\]
for some $B$ of ghost number $-2$, thereby giving $A$ a gauge freedom.
This is an important property to which we shall turn
in the next section.

Having expounded upon the properties of $A$, our next task is clear.
In the following section, we
show that for the tachyon vacuum $\Psi_0$, we can find the
state $A$ satisfying (\ref{condition}) in the approximation of the
level truncation scheme.
\section{Finding The State $A$}
Let us now solve (\ref{condition}) by level truncation.
To do so, let us proceed in two ways. We recall from the previous
section that 
$A$ is well-defined up to the gauge transformation
$A\rightarrow A+Q_{\Psi_0} B$ where $B$ is a state of ghost number
$-2$. 
Because in the level truncation scheme, this gauge invariance is
broken, we first try to find the best fit results without
fixing the gauge of $A$. The fitting procedure is
analogous to that used in \cite{0103103} and we shall not delve too
much into the details. We shall see below that
at level 9, the result is
accurate to $2.4 \%$. However, when we check the behaviour of the
numerical coefficients of $A$ as we increase the accuracy from
level 3 to 9, we found that they do not seem to converge.
We shall explain this phenomenon as the consequence 
of the gauge freedom in the definition of $A$; we shall then redo the
fitting in the
Feynman-Siegel gauge. 
With this second method, we shall find that the coefficients
do converge and the best fit at level 9 is to
$3.2 \%$ accuracy. These results support strongly the existence of
a state $A$ in (\ref{condition}) and hence the statement that the cohomology
around the tachyon vacuum is indeed trivial.
In the following subsections let us present our methods and results in detail.
\subsection{The Fitting without Gauge Fixing $A$}
To solve the condition (\ref{condition}), we first need an explicit 
expression of the identity ${\cal I}$.
Such an expression has been presented in
\cite{gross-jevicki} and \cite{0006240}, differing by a mere
normalization factor 
$-4i$. In this paper, we will follow
the conventions of \cite{0006240} which 
has\footnote{With the normalization $\langle c_1, c_1, c_1 \rangle = 3$ 
that we are
using, we should scale this expression by a factor of $K^3 / 3$, where
$K = 3 \sqrt{3} / 4$. However, as the normalization of the identity
will not change our analysis, we will use this right normalization only
in Section 4, where we are dealing with expressions like ${\cal I}
\star \Phi$.}
\begin{eqnarray}
\label{identity}
\ket{{\cal I}} & = & e^{L_{-2}-\frac{1}{2}L_{-4}+\frac{1}{2} L_{-6}
               -\frac{7}{12} L_{-8}+\frac{2}{3} L_{-10}+...} \ket{0} \\
& = & \ket{0}+L_{-2} \ket{0}+\frac{1}{2}(L_{-2}^2 -L_{-4})\ket{0} \nonumber \\
& + & \left(\frac{1}{6}L_{-2}^3-\frac{1}{4} L_{-2} L_{-4}
-\frac{1}{4} L_{-4} L_{-2}+\frac{1}{2} L_{-6}\right) \ket{0}\nonumber \\
& + & \left(\frac{1}{24} L_{-2}^4+\frac{1}{4}( L_{-2}L_{-6}+L_{-6} L_{-2})
+\frac{1}{8} L_{-4}^2-\frac{7}{12} L_{-8} \right.\nonumber   \\
& & \left.-\frac{1}{12}(L_{-2}^2 L_{-4}+L_{-2} L_{-4} L_{-2}+L_{-4}
L_{-2}^2)\right) \ket{0}\end{eqnarray} 
where $L_n=L_n^m+L_n^g$, the sum of the ghost ($L_n^g$) and matter
($L_n^m$) parts, is the total Virasoro operator.
For later usage we have expanded the exponential up to level 9.
Furthermore, we split $L_n$ into matter and ghost parts and
expand the latter into $b_n,c_n$ operators as\\
$L_m^g := \sum\limits_{n=-\infty}^{\infty} (2m-n) :b_n c_{m-n}: -      
\delta_{m,0}$.
In other words, we write the states in the so-called ``Universal
Basis'' \cite{0006240}.

As a by-product, we have found an elegant expression
for ${\cal I}$ which avoids the recursions\footnote{Indeed the
	expression given in \cite{gj2} has no recursion either,
	however their oscillator expansion is not normal-ordered due
	to ghost insertions at the string mid-point.}
needed to generate the
coefficients in the exponent. In fact, one can show that only $L_{-m}$
for $m$ being a power of 2 survive in the final expression, thus
significantly reducing the complexity of the computation of
level-truncation for ${\cal I}$:
\begin{eqnarray}
\label{IdId}
\ket{\cal{I}} &=& 
  \left( \prod\limits_{n=2}^{\infty} 
	\exp\left\{- \frac{2}{2^n} L_{-2^n}\right\} \right)
  e^{L_{-2}} \ket{0} \nonumber \\
&=& 
\ldots
\exp(-\frac{2}{2^3} L_{-2^3}) \exp(-\frac{2}{2^2} L_{-2^2})
  \exp(L_{-2}) \ket{0},
\end{eqnarray}
where we emphasize that the Virasoro's of higher index stack to the
left {\it ad infinitum}.
We leave the proof of this fact to the Appendix.

It is worth noticing that in the expansion of ${\cal I}$ only odd levels
have nonzero coefficients.
This means that we can constrain the solution $A$ of (\ref{condition}),
if it exists, to have only odd levels in its expansion.
The reason for this is as follows. Equation (\ref{condition}) states that
$Q_B A + \Psi_0 \star A+A\star \Psi_0 = {\cal I}$, moreover we recall
that (cf. e.g. Appendix A.4 of \cite{0103103}) the coefficient
$k_{\ell,i}$ in the expansion of the star product $x \star y =
\sum\limits_{\ell,i} k_{\ell,i} \psi_{\ell,i}$ is
$k_{\ell,i} = \langle \tilde{\psi}_{\ell,i}, x, y \rangle$ for the
orthogonal basis $\tilde{\psi}$ to $\psi$. Now the triple correlator
has the symmetry $\langle x,y,z
\rangle=(-)^{1+g(x)g(y)+\ell(x)+\ell(y)+\ell(z)} \langle x,z,y \rangle$,
where $g(x)$ and $\ell(x)$ are the ghost number and level of the field $x$
respectively. Whence, one can see that the even levels
of $\Psi_0 \star A+A\star \Psi_0$   
will be zero because the tachyon vacuum $\Psi_0 $ has only even
levels and $A$ is constrained to odd levels. 
Furthermore, $Q_B = \sum\limits_n c_n L^m_{-n} + \frac12(m-n):c_m
c_n b_{-m-n}: - c_0$ preserves level. Therefore, in order that both
the left and right hand sides of (\ref{condition})
have only odd levels, $A$ must also have only odd level fields.

Now the procedure is clear. We expand $A$ into odd levels of
ghost number $-1$ with coefficients as parameters and calculate
$Q_{\Psi_0} A$. Indeed as with \cite{0103103}, all the states will be
written as Euclidean vectors whose basis is prescribed by the fields at a
given level; the components of the vectors are thus the expansion
coefficients in each level.
Then we compare $Q_{\Psi_0} A$ with ${\cal I}$ up to the
same level and determine the coefficients of $A$ by minimizing the quantity
$$
\epsilon =\frac{|Q_{\Psi_0} A- {\cal I}|}{|{\cal I}|},
$$
which we of course wish to be as close to zero as possible.
We refer to this as the {\em ``fitting of the coefficients''}.
The norm $|.|$ is the Euclidean norm (for our basis, see the Appendix) .
As observed in \cite{0103085}, different normalizations do not significantly
change the values from the fitting procedure, 
so for simplicity we use the Euclidean norm to define the
above measure of proximity $\epsilon$. The minimum level of the ghost number
$-1$ field $A$ is 3, so we start our fitting from this level and
continue to up to level 9 (higher levels will become computationally
prohibitive). 

First we list the number of components of odd levels for
the fields $A$ and ${\cal I}$ up to given levels:
\[
\begin{array}{|r|c|c|c|c|}  \hline
& $level~$ 3  & $level~$ 5  & $level~$ 7 & $level~$ 9  \\  \hline
\mbox{Number of Components of~}A  &   1 & 4 & 14 & 43  \\ \hline
\mbox{Number of Components of~}{\cal I} & 4 & 14 & 43 & 118 \\ \hline
\end{array}
\] 
From this table, we see that at level 3, we have only one parameter
to fit 4 components. At level 5, we have 4 parameters to fit
14 components. As the level is increased the number of components to 
be fitted increases faster that the number of free parameters. 
Therefore it is not a trivial fitting at all.
\subsubsection{$A$ up to level 3}
At level 3 the identity is:
\[
\begin{array}{rcl}
{\cal I}_3 & = & \ket{0} + L_{-2} \ket{0}\\
& = & \ket{0} - b_{-3}c_1 \ket{0} -2 b_{-2} c_0 \ket{0} + L_{-2}^m \ket{0}
\end{array}
\]
and we find the best fit of $A$ (recall that at level 3 we have only 1
degree of freedom) to be
\[
A_3 = 1.12237 \, b_{-2} \ket{0},
\]
with an $\epsilon$ of $17.1\%$.
\subsubsection{$A$ up to level 5}
Continuing to level 5, we have
\[
\begin{array}{rcl}
{\cal I}_5 & = & \ket{0} + L_{-2} \ket{0} + 
	{1 \over 2}(L_{-2}^2 - L_{-4})\ket{0}\\
& = & \ket{0} - b_{-3}c_1 \ket{0} -2 b_{-2} c_0 \ket{0} + L_{-2}^m
	\ket{0} + b_{-5} c_1 \ket{0} - b_{-2} c_{-2} \ket{0} \\
&&	+ b_{-3} c_{-1} \ket{0} + 2 b_{-3} b_{-2} c_0 c_1 \ket{0} +
	2 b_{-4} c_0 \ket{0} - {1 \over 2} L_{-4}^m \ket{0} \\
&&	- b_{-3}c_1 L_{-2}^m \ket{0} - 2 b_{-2} c_0 L_{-2}^m \ket{0} +
	{1 \over 2} L_{-2}^m L_{-2}^m  \ket{0}.
\end{array}
\]
To this level we have determined the best-fit $A$ to be
\[
A_5 = 1.01893 \, b_{-2} \ket{0} + 0.50921 \, b_{-3} b_{-2} c_1 \ket{0} 
	- 0.518516 \, b_{-4} \ket{0} + 0.504193 \, b_{-2} L_{-2}^m \ket{0},
\]
with an $\epsilon$ of $11.8\%$.

The detailed data of the field $A$ to levels 7 and 9 are given in
table \ref{table_A} of the Appendix. Here we just summarize the
results of the best-fit measure $\epsilon$:
\[
\begin{array}
{|c|r|r|r|r|}  \hline
  & $level~$ 3  & $level~$ 5  & $level~$ 7 & $level~$ 9  \\  \hline
\epsilon=\frac{|Q_{\Psi_0}A-{\cal I}|}{|{\cal I}|} 
	& 0.171484 &  0.117676 & 0.0453748 &0.0243515
\\ \hline
\end{array}
\]
This indicates that up to an accuracy of $2.4 \%$ at level 9, there exists an
$A$ that satisfies (\ref{condition}); moreover the accuracy clearly
gets better with increasing levels. 
This is truly an encouraging result.
\subsection{The Stability of Fitting}
There is a problem however.
Looking carefully at the coefficients of A given in the table 
\ref{table_A}, especially the fitting
coefficients between levels 7 and 9, we see that these
two groups of data have a large difference. Naively it means that our
solution for $A$ does not converge as we increase level.
How do we solve this puzzle?

We recall that $A$ is well-defined only up to the gauge freedom 
$$
A\longrightarrow A+ Q_{\Psi_0} B.
$$
It means that the solutions of (\ref{condition}) should consist of a family
of gauge equivalent $A$.
However, because $Q_{\Psi_0}^2 \neq 0$ under the level truncation
approximation, the family (or the moduli space) is broken into
isolated pieces. Similar phenomena were found in \cite{0103085}
where the momentum-dependent closed states were given by points instead of
a continuous family. Using this fact, our explanation is that
the fitting of levels 7 and 9 are related by 
$Q_{\Psi_0} B$ for some field $B$ of ghost number $-2$. To show this,
we solve a new $\tilde{A}$ up to level 9 that minimizes
$$
\frac{|(\tilde{A})_7-A_7|}{|A_7|}+
\frac{|Q_{\Psi_0}\tilde{A}-{\cal I}_9|}{|{\cal I}_9|}
$$
where $A_7$ is the known fitting data at level seven, ${\cal I}_9$ is the 
identity up to level nine and $(\tilde{A})_7$ refers to
the first 14 components (i.e., the components up to level seven)
of the level 9 expansion of $\tilde{A}$.
By minimizing this above quantity, we balance the stability of fitting
from level 7 to 9.
The data is given in the last column of \ref{table_A}.
Though having gained stability,
the fitting for level 9
is a little worse, with $\epsilon$ increasing from $2.44\%$ to $3.56\%$.

The next thing is to check whether $\tilde{A}-A_9$ is an exact state
$Q_{\Psi_0} B$. We find that this is indeed true and we find a state 
$B$ such that
$$
\frac{|(\tilde{A}-A_9)-Q_{\Psi_0} B|}{|\tilde{A}-A_9|}= 0.28\%.
$$
\subsection{Fitting $A$ in the Feynman-Siegel Gauge}
Alternatively, by gauge-fixing, we can also avoid the instability of
the fit.
If we require the state $A$ to be in the
Feynman-Siegel gauge, $A$ will not have the gauge freedom anymore
and the fitting result should converge as we do not have isolated
points in the gauge moduli space to jump to. We have done so and
do find much greater stability of the coefficients.

Notice that in the Feynman-Siegel gauge, $A$ has the same field bases in
levels 3 and 5, so the fitting at these two levels is the same as in
Subsections 3.1.1 and 3.1.2. However, in this gauge it has one
parameter less at level 7 and 5 less in level 9.
Performing the fit with these parameters we have reached an accuracy
of $\epsilon = 4.8\%$ at level 7 and
$\epsilon = 3.2\%$ at level
9, which is still a good result. The details are presented in
Table \ref{table_A_new} in the Appendix.
\section{Some Subtleties of the Identity}
As pointed out in the Introduction, there are some mysterious and anomalous
features of the identity ${\cal I}$. For example, ${\cal I}$ is not a
normalizable state \cite{HS}, moreover, $c_0$, contrary to expectation,
does not annihilate ${\cal I}$ even though it is a derivation
\cite{0006240}. We shall show in the following that with a slight
modification of the level truncation scheme, this unnormalizability
does not effect the results and furthermore that in our approximation
$Q_{\Psi_0} {\cal I}$ indeed vanishes as it must for consistency.

Let us first show how problems may arise in a naive attempt at level
truncation. Consider the quantity ${\cal I}_{\ell} \star \ket{\Omega}
- \ket{\Omega}$, where ${\cal I}_{\ell}$ denotes the identity
truncated to level $\ell$ and $\ket{\Omega} := c_1 \ket{0}$.
 We of course expect this to approach 0 as
we increase $\ell$. Using the methods of the previous section, we
shall define the measure of proximity
$$
\eta \equiv 
\frac{|{\cal I}_{\ell} \star \ket{\Omega} -\ket{\Omega}|}{|\ket{\Omega}|}
= |{\cal I}_{\ell} \star \ket{\Omega} -\ket{\Omega}|,
$$ 
where $|.|$ is our
usual
norm. We list $\eta$ to levels 3, 5, 7, and 9 in the following Table:
\[
\begin{array}{|c|c|c|c|c|}  \hline
$level~$ \ell	&3 	& 5	&	7	&9 \\ \hline
\eta = |{\cal I}_{\ell} \star \ket{\Omega} - \ket{\Omega}| &
2.06852 & 2.87917 & 3.56054 & 3.9452  \\ \hline
\end{array}
\]
Our $\eta$ obviously does not converge to zero, hence star products
involving ${\cal I}$ do not converge in the usual sense of level
truncation. It is
however not yet necessary to despair, as weak convergence
will come to our rescue\footnote{We thank B. Zwiebach for this
suggestion.}. 

Indeed, instead of
truncating the result to level $\ell$, let us use a slightly different
scheme. We truncate ${\cal I}_{\ell} \star \ket{\Omega}$ to a fixed level $m <
\ell$ and observe how the coefficients of the fields up to level $m$
converge as we increase $\ell$.
In the following table we list the
values of the coefficients coeff($x$) of the 
basis for $m = 2$ (i.e., fields $x$ of level 0, 1 and 2) 
for the expression ${\cal I}_{\ell} \star \ket{\Omega}$.
\[
\begin{array}{|c|r|r|r|r|r|}  \hline
{\cal I}_{\ell} \star \ket{\Omega} 
& $coeff$(\ket{\Omega}) & $coeff$(b_{-1} c_0 \ket{\Omega}) &
$coeff$(b_{-1} c_{-1} \ket{\Omega}) & $coeff$(b_{-2} c_0 \ket{\Omega}) &
$coeff$(L_{-2}^m \ket{\Omega}) \\  \hline
\ell = 3 & $0.6875$ & $0.505181$ &  $-0.905093$ &  $-0.930556$ & $0.465278$  \\
\hline
\ell = 5 & $1.16898$ & $-0.278874$ & $0.38846$ & $0.520748$ & $-0.260374$  \\
\hline
\ell = 7 & $0.911094$ & $0.16252$ & $-0.197833$ & $-0.296607$ & $0.148304$  \\
\hline
\ell = 9 & $1.05767$ & $-0.0971502$ & $0.0902728$ & $0.163579$ & $-0.0817895$  \\
\hline
\end{array}
\]
We see that the $\ket{\Omega}$ component converges to $1$ while the others
converge to $0$, as was hoped. We note however that this (oscillating)
convergence is rather slow and we thus expect slow weak convergence for
other calculations involving the identity.

Having shown that as $\ell \to \infty$  we get a weak convergence
${\cal I}_{\ell} \star \ket{\Omega} \to \ket{\Omega}$, we now consider 
$Q_{\Psi_0} {\cal I}_{\ell}$ as $\ell\to \infty$, which should
tend to zero.
Since $Q_B$ preserves level and
$Q_B {\cal I} = 0$, we have that $Q_B {\cal I} = 0$ in the level
expansion;
thus $Q_{\Psi_0} {\cal I} = \Psi_0 \star {\cal I} - {\cal I} \star
\Psi_0$, which should converge to zero. 

\begin{figure}[!ht]
\leavevmode
\begin{center}
\epsfbox{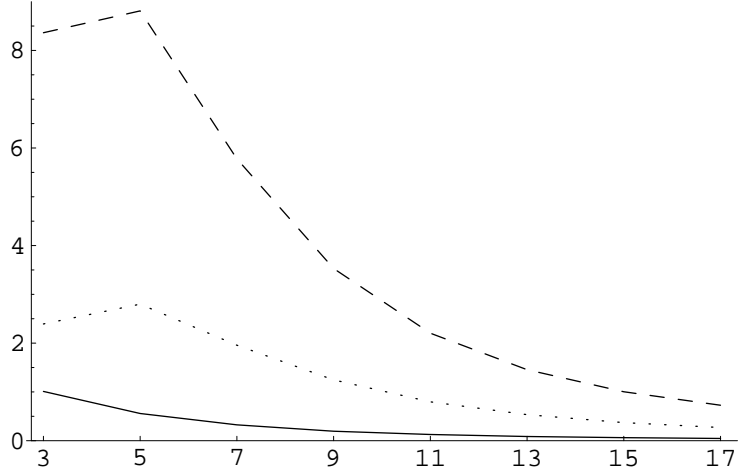}
\end{center}
\caption{A plot of $q_{0,1}(\ell)$ (solid curve), $q_{2,1}(\ell)$
(dotted curve) and $q_{2,3} (\ell)$ (dashed curve) as functions of the
level $\ell$ of the identity. $\ell$ goes from $3$ to $17$.}
\label{conv}
\end{figure}

As the expression $Q_{\Psi_0} {\cal I}$ is
linear in every component of ${\Psi_0}$, that ${\cal I}$ is
$Q_{\Psi_0}$-closed will be
established if we can show that for each component $\phi$ in  ${\Psi_0}$, 
$\phi \star {\cal I} - {\cal I} \star \phi \equiv [\phi \star, {\cal I}]$
converges to zero as the level of ${\cal I}$ is increased. 
We plot in
Fig.\ref{conv}, the absolute values of the coefficient of $c_0 \ket{0}$
in the expressions $\left[(c_1\ket{0}) \star, {\cal I}_{\ell} \right]$, 
$\left[(c_{-1}\ket{0}) \star,
{\cal I}_{\ell} \right]$ and
$\left[(L_{-2}^m c_{1})\ket{0} \star, {\cal I}_{\ell} \right]$,
which we denote by $q_{0,1}(\ell)$,
$q_{2,1}(\ell)$ and $q_{2,3} (\ell)$ respectively. It seems
clear that the coefficients do converge to zero.

The weak convergence we have shown above can be interpreted in a more
abstract setting.
Let us examine the quantity $|{\cal I}_{\ell} \star \Phi - \Phi|$. 
It was
shown in \cite{Wiesbrock} that the $\star$-algebra of the open bosonic
string field theory is a $C^*$-algebra. A well-known theorem dictates that
any $C^*$-algebra $M$ (with or without unit) has a so-called {\em
approximate identity} which is a set of operators $\left\{ {\cal I}_i
\right\}$ in $M$ indexed by $i$ satisfying
(i) $\| {\cal I}_i \| \le 1$ for every $i$ and (ii) $\|{\cal I}_i x - x\|
\rightarrow 0$ and $\|x {\cal I}_i - x\| \rightarrow 0$ for all $x \in
M$ with respect to the (Banach) norm $\|.\|$ of $M$ (cf. e.g. \cite{Zhu}).

The level $\ell$ in our level truncation scheme is suggestive of an 
index for ${\cal I}$. Furthermore the weak convergence we have found in
this section is analogous to property (ii) of the theorem 
(being of course a little cavalier about the distinction of the Banach
norm of the $C^*$-algebra with the Euclidean norm used here). 
Barring this
subtlety, it is highly suggestive that our ${\cal I}_{\ell}$ is an
approximate identity of the $\star$-algebra indexed by level $\ell$.
\section{Conclusion and Discussions}
According to a strong version of Sen's Second Conjecture, there should
be an absence of any open string states around the perturbatively
stable tachyon vacuum $\Psi_0$. This disappearance of all states, not
merely the physical ones of ghost number 1, means that the cohomology of the
new BRST operator $Q_{\Psi_0}$ should be completely trivial near the
vacuum. It is the key observation of this paper that this statement of
triviality is implied by the existence of
a ghost number $-1$ field $A$ satisfying 
$$
Q_{\Psi_0} A=Q_B A+\Psi_0 \star A+ A\star \Psi_0={\cal I}.
$$
That is to say that if the identity of the $\star$-algebra ${\cal I}$
is a $Q_{\Psi_0}$ exact state, then the cohomology of
$Q_{\Psi_0}$ would be trivial.

The level truncation scheme was subsequently applied to 
check our proposal. We have found that such a state $A$ exists 
up to an accuracy of $3.2 \%$ at level 9.
Although these numerical results give a strong support to the proposal
for the existence of $A$ and hence the triviality of
$Q_{\Psi_0}$-cohomology near the vacuum,
an analytic expression for $A$ would be most welcome. 
However, to obtain such an analytic form of $A$, it seems that we would
require the analytic expression for the vacuum
$\Psi_0$, bringing us back to an old problem.
It is perhaps possible that by choosing different
gauges other than the Feynman-Siegel gauge we may find such a solution.

Our solution $A$ satifies  $\{A,Q\}=I$. 
It would be nice to see whether we can choose 
$A$ cleverly to make $A\star A=0$ (our Feynman-Siegel gauge fitting
may not satisfy this equation). We are interested with this case because
for the proposal of $Q_B = c_n + (-)^n c_{-n}$ made in
\cite{0012251,0102112}, 
one could find that $A = \frac12(b_{-n} + (-)^n b_{n})$ which does
satisfy $A^2 = 0$. It would be interesting to mimic this nilpotency
within the $\star$-algebra. 
Furthermore, it would be fascinating 
to see if we can make a field redefinition to 
reduce $A$ to a simple operator such as $b_0$, and at the same
time reduce $Q_{\Psi_0}$ to a new BRST operator as
suggested in \cite{0102112}, for example, $c_0$.

Last but not least, an interesting question is about the identity
${\cal I}$. In this paper we have given an elegant analytic expression
for ${\cal I}$ which avoids the usage of complicated recursion
relations. Furthermore, we have suggested that though the $\star$-algebra
of OSFT may be a non-unital $C^*$-algebra, ${\cal I}$ still may serve
as a so-called approximate identity.
However,
as we discussed before, anomalies related to the identity in the String Field
Theory make the calculation in level truncation converge 
very slowly. It will be useful to understand more about
${\cal I}$.

\section*{Acknowledgements}
We would like to express our sincere gratitude to B.~Zwiebach for many
insightful comments and useful discussions as well as careful
proof-reading of the manuscript.
Also we are indebted to
H.~Hata, A.~Sen and W.~Taylor for kind remarks and suggestions on the
preliminary draft of this paper.
\appendix
\section{The Perturbatively Stable Vacuum Solution at Level
$(M,3M)$}
We tabulate the coefficient of the expansion of the stable vacuum solution
$\Psi_0$ at various levels and interaction \cite{0002237}.
\[
{\tiny
\begin{tabular}{|l|l|l|l|l|}  \hline
$gh = 1$ field basis &  level $(2,6)$  & level  $(4,12)$ & level $(6,18)$  & 
level $(8,24)$  
\\  \hline \hline
$\ket{\Omega}$   &   $0.3976548947184288$  & $0.4007200390749924$ &
$0.4003790755638671$  & $0.39973608190423154$
\\  \hline \hline
$b_{-1} c_{-1}\ket{\Omega}$   & $-0.1389738152295008$  & 
$-0.1502869559917484$ & $-0.15477497270540513$ & $-0.15712091953765914$
\\ \hline
$L_{-2}^m \ket{\Omega}$   & $ 0.0408931493261807$ & $0.04159452148973691$ &
$0.04175525359702033$ & $0.041806849347695574$
\\ \hline  \hline
$b_{-1} c_{-3}\ket{\Omega}$ &  & $0.041073385934010505$ &
$0.041936906548529496$ & $0.042358626301118626$   
\\ \hline
$b_{-2} c_{-2}\ket{\Omega}$ &  &  $ 0.02419174563180113$ 
&  $0.02489022878379843 $  & $0.025301843897808124 $ 
\\ \hline
$b_{-3} c_{-1}\ket{\Omega}$ &  &$ 0.013691128644670262$
&  $0.013978968849509828 $  & $0.014119542100372846 $
\\ \hline 
$L_{-4}^m \ket{\Omega}$ &  &  $-0.003741923212578628$
&  $-0.0037331617302832193 $  & $ -0.0037279001402683682 $
\\ \hline 
$b_{-1} c_{-1}L_{-2}^m \ket{\Omega}$ & & $0.005013189182427192$
&  $0.005410660944694899 $  & $0.005620705137023851 $
\\ \hline
$L_{-2}^m L_{-2}^m \ket{\Omega}$ &  & $-0.00043064009114185083$
&  $-0.0004545462255696699 $  & $-0.0004654022166127481 $
\\ \hline  \hline 
$b_{-1}c_{-5}\ket{\Omega}$ &  & & $-0.02193107815206234$ & 
$-0.022161386573208323 $
\\ \hline
$ b_{-2} c_{-4}\ket{\Omega}$ &  & & 
$-0.013702048066242712 $ & $-0.01385275004340868 $  
\\ \hline
$ b_{-3}  c_{-3}\ket{\Omega}$ &  & &
$-0.00834273227278023 $ & $-0.008359650003474304 $
\\ \hline 
$ b_{-4}  c_{-2}\ket{\Omega}$ &  & &
$-0.0068510240331213544 $ & $-0.0069263750217043295 $
\\ \hline  
$ b_{-5}  c_{-1}\ket{\Omega}$ &  & &
$-0.004386215630412471 $ & $-0.0044322773146416965 $
\\ \hline    
$ b_{-2} b_{-1} c_{-2} c_{-1}\ket{\Omega}$ &  & &
$-0.005651485281802872 $ & $-0.00580655453652034 $
\\ \hline    
$ L^m_{-6}\ket{\Omega}$ &  & &
$0.0010658398347450269 $ & $0.0010617366766707361 $
\\ \hline 
$ b_{-1}c_{-1}  L^m_{-4}\ket{\Omega}$ &  & &
$-0.0008498595740547494 $ & $-0.0008732233330861659 $
\\ \hline 
$ b_{-1}c_{-2}  L^m_{-3}\ket{\Omega}$ &  & &
$-0.000046769138331183204 $ & $-0.000052284121618944255 $
\\ \hline 
$b_{-2}  c_{-1} L^m_{-3}\ket{\Omega}$ &  & &
$-0.000023384569165591568 $ & $-0.000026142060809472097 $
\\ \hline 
$ L^m_{-3}  L^m_{-3}\ket{\Omega}$ &  & &
$4.479437511126653\times 10^{-6} $ & $5.080488681869039\times 10^{-6} $
\\ \hline 
$ b_{-1}c_{-3}  L^m_{-2}\ket{\Omega}$ &  & &
$-0.002457790374962076 $ & $-0.002528657337188949 $
\\ \hline 
$b_{-2}  c_{-2}   L^m_{-2}\ket{\Omega}$ &  & &
$-0.0020680241879350277 $ & $-0.002125416342663475 $
\\ \hline 
$ b_{-3}  c_{-1}  L^m_{-2}\ket{\Omega}$ &  & &
$-0.0008192634583206926 $ & $-0.0008428857790629816 $
\\ \hline 
$  L^m_{-4}  L^m_{-2}\ket{\Omega}$ &  & &
$0.00022330350231085353 $ & $0.00022500193649010967 $
\\ \hline 
$b_{-1}c_{-1}  L^m_{-2}  L^m_{-2}\ket{\Omega}$ &  & &
$-0.00011131535311028013 $ & $-0.00012817322136544294 $
\\ \hline 
$  L^m_{-2}  L^m_{-2}  L^m_{-2}\ket{\Omega}$ &  & &
$-7.241008154399294\times 10^{-6} $ & $-6.240064701718801\times 10^{-6} $
\\ \hline
\end{tabular}
}
\]
\begin{flushright}{\it continued...}\end{flushright}

\be
\label{table_psi}
{\tiny
\begin{tabular}{|l|l|l|l|l|}  \hline
$gh = 1$ field basis &  level $(2,6)$  & level  $(4,12)$ & level $(6,18)$  & 
level $(8,24)$  
\\  \hline \hline
$b_{-1}c_{-7}\ket{\Omega}$ &  & & & $0.014312021693536028 $  
\\ \hline
$b_{-2}c_{-6}\ket{\Omega}$ &  & & & $0.009158200585940239 $  
\\ \hline
$ b_{-3} c_{-5}\ket{\Omega}$ &  & & & $0.005674268936470511 $  
\\ \hline
$b_{-4}  c_{-4}\ket{\Omega}$ &  & & & $0.004838957768226669 $  
\\ \hline
$ b_{-5}  c_{-3}\ket{\Omega}$ &  & & & $0.0034045613618823045 $  
\\ \hline
$b_{-6}  c_{-2}\ket{\Omega}$ &  & & & $0.0030527335286467446 $  
\\ \hline
$b_{-2} b_{-1} c_{-3}  c_{-2}\ket{\Omega}$ &  & & & $-0.0035422558218537676 $  
\\ \hline
$b_{-7}  c_{-1}\ket{\Omega}$ &  & & & $0.0020445745276480116 $  
\\ \hline
$b_{-2} b_{-1} c_{-4}  c_{-1}\ket{\Omega}$ &  & & & $ 0.0037527555019998804 $  
\\ \hline
$b_{-3} b_{-1} c_{-3}  c_{-1}\ket{\Omega}$ &  & & & $0.0004302428004449616 $  
\\ \hline
$b_{-3}  b_{-2}  c_{-2}  c_{-1} \ket{\Omega}$ &  & & & $-0.0011807519406179202 $  
\\ \hline
$b_{-4}  b_{-1} c_{-2}  c_{-1}\ket{\Omega}$ &  & & & $0.0018763777509999383 $  
\\ \hline
$  L^m_{-8}\ket{\Omega}$ &  & & & $ -0.00041801038699211334 $  
\\ \hline
$b_{-1}c_{-1}  L^m_{-6}\ket{\Omega}$ &  & & & $0.00029329813765991303 $  
\\ \hline
$b_{-1}c_{-2}  L^m_{-5}\ket{\Omega}$ &  & & & $6.281489731737461\times 10^{-6} $  
\\ \hline
$ b_{-2}  c_{-1}   L^m_{-5} \ket{\Omega}$ &  & & & $3.140744865868727\times 10^{-6} $  
\\ \hline
$b_{-1}c_{-3}  L^m_{-4}\ket{\Omega}$ &  & & & $0.000500528172313894 $  
\\ \hline
$b_{-2}  c_{-2}  L^m_{-4}\ket{\Omega}$ &  & & & $0.00030379159554779373 $  
\\ \hline
$ b_{-3}  c_{-1}  L^m_{-4}\ket{\Omega}$ &  & & & $0.00016684272410463048 $  
\\ \hline
$ L^m_{-4}   L^m_{-4}\ket{\Omega}$ &  & & & $-0.000021999720024591806 $  
\\ \hline
$ b_{-1}c_{-4}  L^m_{-3}\ket{\Omega}$ &  & & & $0.00003496149452657495 $  
\\ \hline
$ b_{-2}c_{-3}  L^m_{-3}\ket{\Omega}$ &  & & & $-3.2753561169368668\times 10^{-6} $  
\\ \hline
$b_{-3}  c_{-2}  L^m_{-3}\ket{\Omega}$ &  & & & $-2.1835707446245427\times 10^{-6} $  
\\ \hline
$ b_{-4}  c_{-1}   L^m_{-3}\ket{\Omega}$ &  & & & $8.74037363164371\times 10^{-6} $  
\\ \hline
$  L^m_{-5}   L^m_{-3} \ket{\Omega}$ &  & & & $-1.3196771313891132\times 10^{-6} $  
\\ \hline
$b_{-1}c_{-1}  L^m_{-3}  L^m_{-3} \ket{\Omega}$ &  & & & $1.2594432286572633\times 10^{-6} $  
\\ \hline
$b_{-1}c_{-5}  L^m_{-2}\ket{\Omega}$ &  & & & $0.001534533432927412 $  
\\ \hline
$ b_{-2}c_{-4}   L^m_{-2}\ket{\Omega}$ &  & & & $0.0013556709245221895 $  
\\ \hline
$ b_{-3}  c_{-3}   L^m_{-2}\ket{\Omega}$ &  & & & $0.0006166063072874846 $  
\\ \hline
$b_{-4}  c_{-2}  L^m_{-2}\ket{\Omega}$ &  & & & $0.0006778354622610939 $  
\\ \hline
$ b_{-5}  c_{-1}   L^m_{-2}\ket{\Omega}$ &  & & & $0.00030690668658548353 $  
\\ \hline
$ b_{-2} b_{-1} c_{-2}  c_{-1}  L^m_{-2}\ket{\Omega}$ &  & & & $0.0005782814358972997 $  
\\ \hline
$  L^m_{-6}   L^m_{-2} \ket{\Omega}$ &  & & & $-0.00007624602726052426 $  
\\ \hline
$b_{-1}c_{-1}  L^m_{-4}  L^m_{-2}\ket{\Omega}$ &  & & & $0.00006375616369006518 $  
\\ \hline
$ b_{-1}c_{-2}  L^m_{-3}  L^m_{-2}\ket{\Omega}$ &  & & & $5.9626436110722614\times 10^{-6} $  
\\ \hline
$b_{-2}  c_{-1}   L^m_{-3}  L^m_{-2}\ket{\Omega}$ &  & & & $2.9813218055361256\times 10^{-6} $  
\\ \hline
$ L^m_{-3}  L^m_{-3}  L^m_{-2} \ket{\Omega}$ &  & & & $-5.422796727699355\times 10^{-7} $  
\\ \hline
$b_{-1}c_{-3}  L^m_{-2}  L^m_{-2}\ket{\Omega}$ &  & & & $0.00004728162691342103 $  
\\ \hline
$ b_{-2}  c_{-2}   L^m_{-2}  L^m_{-2}\ket{\Omega}$ &  & & & $0.00010011937816215435 $  
\\ \hline
$ b_{-3}  c_{-1}   L^m_{-2}  L^m_{-2}\ket{\Omega}$ &  & & & $0.000015760542304474034 $  
\\ \hline
$  L^m_{-4}  L^m_{-2}  L^m_{-2}\ket{\Omega}$ &  & & & $-4.371565449219928\times 10^{-6} $  
\\ \hline
$b_{-1}c_{-1}  L^m_{-2}  L^m_{-2}  L^m_{-2}\ket{\Omega}$ &  & & & $-3.759766768481099\times 10^{-7} $  
\\ \hline
$ L^m_{-2}  L^m_{-2}  L^m_{-2}  L^m_{-2}\ket{\Omega}$ &  & & & $7.259081254041818\times 10^{-7} $  
\\ \hline
\end{tabular}
}
\ee
\section{Fitting of the Parameters of $A$}
\subsection{$A$ up to Level 9 without Gauge Fixing}
As $A$ is of ghost number $-1$ and has only odd levels, 
we here tabulate such field basis at levels 3, 5, 7 and 9. 
The best-fit numbers are the 
coefficients of $A$ obtained by best-fit via minimizing
$\epsilon =\frac{|Q_{\Psi_0} A- {\cal I}|}{|{\cal I}|}$.
The stable fit at level 9 is constructed so as to control the
convergence behaviour of the coefficients.
\be
\label{table_A}
{\tiny
\begin{tabular}{|r|r|r|r|r|r|} \hline
Field Basis  &  level 3 fit  &  level 5 fit & level 7 fit  &  level 9 fit 
& stable level 9 fit
\\ \hline
$b_{-2}\ket{0}$  &  $1.12237 $ & $1.01893 $ & $0.948316 $ & $1.25995 $ 
& $0.931864 $
\\ \hline  \hline
$b_{-3}  b_{-2}  c_{1}\ket{0}$ & & $0.50921 $ & $0.37306 $ & $0.660674 $ 
& $0.401547$
\\  \hline
$b_{-4}\ket{0}$ & & $-0.518516 $ & $-0.753272 $ & $-0.25828 $ 
& $-0.753004$
\\ \hline
$b_{-2}  L^m_{-2}\ket{0}$ & & $0.504193 $ & $0.50695 $ & $0.400769 $ 
& $0.496562$
\\ \hline \hline
$b_{-4}  b_{-3}  c_{1}\ket{0}$ & & & $ 0.698601 $ & $-0.10683 $ 
& $0.691255$
\\ \hline
$b_{-5}  b_{-2}  c_{1}\ket{0}$ & & & $0.893251 $ & $-1.8453 $ 
& $0.888407$
\\ \hline
$b_{-6}\ket{0}$ & & & $-0.531323 $ & $ 1.40819 $ 
& $-0.541737$
\\ \hline
$-b_{-3}  b_{-2}  c_{-1}\ket{0}$ & & & $-1.87167 $ & $3.14822 $ 
& $-1.86475$
\\ \hline
$-b_{-4}  b_{-2}  c_{0}\ket{0}$ & & & $-2.54254 $ & $ 3.2966 $ 
& $ -2.54625$
\\ \hline
$b_{-2} L^m_{-4}\ket{0}$ & & & $ 0.264611 $ & $-0.750856 $ 
& $0.255304 $
\\ \hline
$b_{-3}  L^m_{-3}\ket{0}$ & & & $0.00193005 $ & $-0.0539165 $ 
& $-0.0191971$
\\ \hline
$b_{-3}  b_{-2}  c_{1}  L^m_{-2}\ket{0}$ & & & $0.358002 $ & $0.301463 $ 
& $0.338645$
\\ \hline
$b_{-4}  L^m_{-2}\ket{0}$ & & & $-0.724095 $ & $ 0.163428 $ 
& $-0.744985$
\\ \hline
$b_{-2}  L^m_{-2}  L^m_{-2}\ket{0}$ & & & $0.166002 $ & $0.180328 $ 
& $0.169096$
\\ \hline  \hline
$b_{-5}  b_{-4}  c_{1}\ket{0}$ & & & &$0.0796036 $ 
& $ 0.273844$
\\ \hline 
$b_{-6}  b_{-3}  c_{1}\ket{0}$ & & && $-1.09893 $ 
& $-0.107261$
\\ \hline 
$b_{-7}  b_{-2}  c_{1}\ket{0}$ & & && $ 0.847731 $ 
& $0.195816$
\\ \hline 
$b_{-8}\ket{0}$ & & && $-0.313743 $ 
& $-0.277211$
\\ \hline 
$b_{-3}  c_{-3} b_{-2}\ket{0}$ & & && $ -19.0376 $ 
& $ -4.11409$
\\ \hline 
$-b_{-4}  b_{-2}  c_{-2}\ket{0}$ & & && $ -0.147445 $ 
& $-0.626872$
\\ \hline 
$-b_{-4}  b_{-3}  c_{-1}\ket{0}$ & & && $1.80597 $ 
& $-0.0745503$
\\ \hline 
$-b_{-5}  b_{-2}  c_{-1}\ket{0}$ & & && $-0.172462 $ 
& $-0.356920$
\\ \hline 
$b_{-4}  b_{-3}  b_{-2}  c_{0}  c_{1}\ket{0}$ & & && $ 1.05994$ 
& $-0.102556$
\\ \hline 
$-b_{-5}  b_{-3}  c_{0}\ket{0}$ & & & &$1.48397 $ 
& $-0.319450$
\\ \hline 
$ -b_{-6}  b_{-2}  c_{0}\ket{0}$ & & && $ -0.784562 $ 
& $0.0949989$
\\ \hline 
$ b_{-2} L^m_{-6}\ket{0}$ & & && $0.103719 $ 
& $-0.00879977$
\\ \hline 
$b_{-3}  L^m_{-5}\ket{0}$ & & && $-0.530976 $ 
& $ -0.0537990$
\\ \hline 
$ b_{-3}  b_{-2}  c_{1}  L^m_{-4}\ket{0}$ & & & &$ 0.428303 $ 
& $0.0633010$
\\ \hline 
$b_{-4}  L^m_{-4}\ket{0}$ & & && $0.114766 $ 
& $0.111182$
\\ \hline 
$b_{-4}  b_{-2}  c_{1}  L^m_{-3}\ket{0}$ & & && $0.687831 $ 
& $ 0.200100$
\\ \hline 
$b_{-5}  L^m_{-3}\ket{0}$ & & & &$-0.165379 $ 
& $-0.134011$
\\ \hline 
$-b_{-3}  b_{-2}  c_{0}  L^m_{-3}\ket{0}$ & & & &$ -2.72288 $ 
& $-0.722198$
\\ \hline 
$b_{-2}  L^m_{-3} L^m_{-3}\ket{0}$ & & & &$ 0.3427 $ 
& $0.0910701$
\\ \hline 
$b_{-4}  b_{-3}  c_{1}  L^m_{-2}\ket{0}$ & & & &$-0.01845 $ 
& $0.304266$
\\ \hline 
$ b_{-5}  b_{-2}  c_{1}  L^m_{-2}\ket{0}$ & & & &$-0.628564 $ 
& $-0.137309$
\\ \hline 
$b_{-6}  L^m_{-2}\ket{0}$ & & && $ 0.39923 $ 
& $0.195490$
\\ \hline 
$-b_{-3}  b_{-2}  c_{-1}  L^m_{-2}\ket{0}$ & & && $-0.537685 $ 
& $-0.289167$
\\ \hline 
$ -b_{-4}  b_{-2} c_{0}  L^m_{-2}\ket{0}$ & && & $ 0.951973 $ 
& $-0.288878$
\\ \hline 
$ b_{-2}  L^m_{-4}  L^m_{-2}\ket{0}$ & & & &$-0.237783 $ 
& $-0.0856879$
\\ \hline 
$b_{-3}  L^m_{-3}  L^m_{-2}\ket{0}$ & & & &$-0.332135 $ 
& $-0.0868470$
\\ \hline 
$b_{-3}  b_{-2}  c_{1}  L^m_{-2}  L^m_{-2}\ket{0}$ & & & &$ 0.128844 $ 
& $0.126029$
\\ \hline 
$b_{-4}  L^m_{-2}  L^m_{-2}\ket{0}$ & & && $-0.00185911 $ 
& $-0.160345$
\\ \hline 
$b_{-2}  L^m_{-2}  L^m_{-2}  L^m_{-2}\ket{0}$ & & && $ 0.0403381 $ 
& $0.0402361$
\\ \hline  \hline
$\epsilon=|Q_{\Psi_0} A-{\cal I}|/|{\cal I}|$ & $0.171484$ &
$0.117676$ & $0.0453748$ &$0.0243515$ & $0.0356226$
\\ \hline
\end{tabular}
}
\ee
\subsection{Fitting $A$ in the Feynman-Siegel gauge}
As $A$ enjoys the gauge freedom $A\rightarrow A+ Q_{\Psi_0} B$, we can
fix it to be in the Feynman-Siegel gauge. This is another way
to control the convergence behaviour of the coefficients.
\be
\label{table_A_new}
{\tiny
\begin{tabular}{|r|r|r|r|r|} \hline
fields   &  level 3 fit  &  level 5 fit & level 7 fit  &  level 9 fit 
\\ \hline
$b_{-2}\ket{0}$  &  $1.12237 $ & $1.01893 $ & 
$1.12465$  &  $1.05322 $  
\\ \hline  \hline
$b_{-3}  b_{-2}  c_{1}\ket{0}$ & & $0.50921 $ & 
$ 0.467$  &  $0.500266 $ 
\\  \hline
$b_{-4}\ket{0}$ & & $-0.518516 $ & 
$-0.503772$  &  $ -0.53228 $ 
\\ \hline
$b_{-2}  L^m_{-2}\ket{0}$ & & $0.504193 $ & 
$0.476325$  &  $ 0.504269$ 
\\ \hline \hline
$b_{-4}  b_{-3}  c_{1}\ket{0}$ & & & 
$0.333428$  &  $0.326986 $ 
\\ \hline
$b_{-5}  b_{-2}  c_{1}\ket{0}$ & & & 
$-0.330557$  &  $-0.328381 $ 
\\ \hline
$b_{-6}\ket{0}$ & & & 
$0.346811$  &  $0.331188 $ 
\\ \hline
$-b_{-3}  b_{-2}  c_{-1}\ket{0}$ & & & 
$0.325862$  &  $0.327997 $ 
\\ \hline
$-b_{-4}  b_{-2}  c_{0}\ket{0}$ & & & 
$0$  &  $ 0$ 
\\ \hline
$b_{-2} L^m_{-4}\ket{0}$ & & & 
$-0.166799$  &  $-0.164306 $ 
\\ \hline
$b_{-3}  L^m_{-3}\ket{0}$ & & & 
$0.00133026$  &  $0.000334022 $ 
\\ \hline
$b_{-3}  b_{-2}  c_{1}  L^m_{-2}\ket{0}$ & & & 
$0.341592$  &  $0.328637 $ 
\\ \hline
$b_{-4}  L^m_{-2}\ket{0}$ & & & 
$ -0.332864$  &  $-0.327326 $ 
\\ \hline
$b_{-2}  L^m_{-2}  L^m_{-2}\ket{0}$ & & & 
$ 0.1686$  &  $0.165931 $ 
\\ \hline  \hline
$b_{-5}  b_{-4}  c_{1}\ket{0}$ & & & 
$$  &  $0.245489 $ 
\\ \hline 
$b_{-6}  b_{-3}  c_{1}\ket{0}$ & & &
$$  &  $-0.253014 $ 
\\ \hline 
$b_{-7}  b_{-2}  c_{1}\ket{0}$ & & &
$$  &  $0.250149 $ 
\\ \hline 
$b_{-8}\ket{0}$ & & &
$$  &  $-0.257672 $ 
\\ \hline 
$b_{-3}  c_{-3} b_{-2}\ket{0}$ & & &
$$  &  $0.249999 $ 
\\ \hline 
$-b_{-4}  b_{-2}  c_{-2}\ket{0}$ & & &
$$  &  $-0.256812 $ 
\\ \hline 
$-b_{-4}  b_{-3}  c_{-1}\ket{0}$ & & &
$$  &  $0.246526 $ 
\\ \hline 
$-b_{-5}  b_{-2}  c_{-1}\ket{0}$ & & &
$$  &  $-0.25213 $ 
\\ \hline 
$b_{-4}  b_{-3}  b_{-2}  c_{0}  c_{1}\ket{0}$ & & &
$$  &  $0 $ 
\\ \hline 
$-b_{-5}  b_{-3}  c_{0}\ket{0}$ & & & 
$$  &  $0 $ 
\\ \hline 
$ -b_{-6}  b_{-2}  c_{0}\ket{0}$ & & &
$$  &  $0 $ 
\\ \hline 
$ b_{-2} L^m_{-6}\ket{0}$ & & &
$$  &  $0.00104113 $ 
\\ \hline 
$b_{-3}  L^m_{-5}\ket{0}$ & & &
$$  &  $0.0000151443 $ 
\\ \hline 
$ b_{-3}  b_{-2}  c_{1}  L^m_{-4}\ket{0}$ & & & 
$$  &  $-0.126025 $ 
\\ \hline 
$b_{-4}  L^m_{-4}\ket{0}$ & & &
$$  &  $0.12448 $ 
\\ \hline 
$b_{-4}  b_{-2}  c_{1}  L^m_{-3}\ket{0}$ & & &
$$  &  $ -0.0004548$ 
\\ \hline 
$b_{-5}  L^m_{-3}\ket{0}$ & & & 
$$  &  $-0.000819122 $ 
\\ \hline 
$-b_{-3}  b_{-2}  c_{0}  L^m_{-3}\ket{0}$ & & & 
$$  &  $0 $ 
\\ \hline 
$b_{-2}  L^m_{-3} L^m_{-3}\ket{0}$ & & & 
$$  &  $ 0.0000905036$ 
\\ \hline 
$b_{-4}  b_{-3}  c_{1}  L^m_{-2}\ket{0}$ & & & 
$$  &  $0.250728 $ 
\\ \hline 
$ b_{-5}  b_{-2}  c_{1}  L^m_{-2}\ket{0}$ & & & 
$$  &  $-0.251499 $ 
\\ \hline 
$b_{-6}  L^m_{-2}\ket{0}$ & & &
$$  &  $ 0.250865 $ 
\\ \hline 
$-b_{-3}  b_{-2}  c_{-1}  L^m_{-2}\ket{0}$ & & &
$$  &  $  0.249179$ 
\\ \hline 
$ -b_{-4}  b_{-2} c_{0}  L^m_{-2}\ket{0}$ & && 
$$  &  $ 0$ 
\\ \hline 
$ b_{-2}  L^m_{-4}  L^m_{-2}\ket{0}$ & & & 
$$  &  $ -0.123363$ 
\\ \hline 
$b_{-3}  L^m_{-3}  L^m_{-2}\ket{0}$ & & & 
$$  &  $0.000457948 $ 
\\ \hline 
$b_{-3}  b_{-2}  c_{1}  L^m_{-2}  L^m_{-2}\ket{0}$ & & & 
$$  &  $0.126358 $ 
\\ \hline 
$b_{-4}  L^m_{-2}  L^m_{-2}\ket{0}$ & & &
$$  &  $-0.125248 $ 
\\ \hline 
$b_{-2}  L^m_{-2}  L^m_{-2}  L^m_{-2}\ket{0}$ & & &
$$  &  $ 0.0406385$ 
\\ \hline  \hline
$\epsilon=|Q_{\Psi_0} A-{\cal I}|/|{\cal I}|$ & $0.171484$ &  $0.117676$ & 
$0.0480658$  &  $0.0320384 $ 
\\ \hline
\end{tabular}
}
\ee
\subsection{Expansion of ${\cal I}$ up to level 9}
Immediately below the field basis at ghost number 0 and levels 1, 3,
5, 7 and 9 is given the coefficient of the expansion of ${\cal I}$.
\be
\label{table_I}
{\tiny
\begin{tabular}{|r|r|r|r|r|} \hline 
$\ket{0}$ & $b_{-3} c_1\ket{0}$ & $-b_{-2} c_{0} \ket{0}$ & $L^m_{-2}\ket{0}$  &
$b_{-5} c_1\ket{0} $
\\ \hline $1 $   & $-1 $  &$2 $   & $1 $   &  $1 $   \\ \hline \hline  
 $-b_{-2}  c_{-2}\ket{0} $ &  $-b_{-3}  c_{-1}\ket{0} $ 
& $b_{-3} b_{-2} c_{0}c_1 \ket{0} $ &
$-b_{-4}  c_{0}\ket{0} $  & $L^m_{-4}\ket{0} $ 
\\ \hline $1 $   & $-1 $  &$ 2$   & $ -2$ & $-\frac{1}{2} $ \\ \hline \hline 
$b_{-2} c_1 L^m_{-3}\ket{0} $ 
& $b_{-3} c_1 L^m_{-2}\ket{0} $ & $-b_{-2}  c_{0} L^m_{-2}\ket{0} $  
& $L^m_{-2}L^m_{-2}\ket{0} $ &  $b_{-7} c_1\ket{0} $ 
\\ \hline $0 $   & $-1$  &$ 2$   & $\frac{1}{2} $   &  $-1 $   \\ \hline \hline
$-b_{-2} c_{-4}\ket{0} $   & $-b_{-3} c_{-3}\ket{0} $  &$b_{-3} b_{-2} c_{-2}c_1\ket{0} $ 
  & $-b_{-4}  c_{-2}\ket{0} $   &  $b_{-4} b_{-2} c_{-1}c_1\ket{0} $
\\ \hline $ 0$   & $ -1$  &$ 1$   & $0 $   &  $0 $   \\ \hline \hline
$-b_{-5} c_{-1}\ket{0} $   & $b_{-4} b_{-3} c_{0}c_1\ket{0} $  &$b_{-5} b_{-2} c_{0}c_1\ket{0} $
   & $-b_{-6}  c_{0}\ket{0} $   &  $b_{-3} b_{-2}  c_{-1} c_{0}\ket{0} $
\\ \hline $1 $   & $2 $  &$-2 $   & $ 2$   &  $2 $   \\ \hline \hline
 $L^m_{-6}\ket{0} $   & $b_{-2} c_1 L^m_{-5}\ket{0} $  &$b_{-3}c_1 L^m_{-4}\ket{0} $   
& $-b_{-2} c_{0} L^m_{-4}\ket{0} $   &  $b_{-4}c_1 L^m_{-3}\ket{0} $
\\ \hline $0 $   & $0 $  &$1/2 $   & $ -1$   &  $0 $   \\ \hline \hline
$-b_{-2} c_{-1} L^m_{-3}\ket{0} $   & $-b_{-3}  c_{0} L^m_{-3}\ket{0} $  
&$L^m_{-3}L^m_{-3}\ket{0} $   & $b_{-5} c_1 L^m_{-2}\ket{0} $   & 
 $-b_{-2}  c_{-2} L^m_{-2}\ket{0} $ 
\\ \hline $ 0$   & $ 0$  &$0 $   & $1 $   &  $1 $   \\ \hline \hline
$-b_{-3}  c_{-1} L^m_{-2}\ket{0} $   & $b_{-3} b_{-2} c_{0}c_1 L^m_{-2}\ket{0} $  
&$-b_{-4}  c_{0} L^m_{-2}\ket{0} $   & $L^m_{-4} L^m_{-2}\ket{0} $   &  
$b_{-2} c_1 L^m_{-3} L^m_{-2}\ket{0} $ 
\\ \hline $ -1$   & $2 $  &$-2 $   & $-1/2 $   &  $ 0$   \\ \hline \hline
$b_{-3}c_{1}L^m_{-2}L^m_{-2}\ket{0} $   & $-b_{-2}c_{0}L^m_{-2}L^m_{-2}\ket{0} $  
&$L^m_{-2}L^m_{-2}L^m_{-2}\ket{0} $   & $b_{-9}c_1\ket{0} $   &  $-b_{-2} c_{-6}\ket{0} $ 
\\ \hline $-1/2 $   & $ 1$  &$1/6 $   & $ 1$   &  $0 $   \\ \hline \hline
$-b_{-3} c_{-5}\ket{0} $   & $b_{-3} b_{-2} c_{-4}c_1\ket{0} $  
&$-b_{-4} c_{-4}\ket{0} $   & $b_{-4} b_{-2} c_{-3}c_1\ket{0} $   & 
 $-b_{-5}  c_{-3}\ket{0} $ 
\\ \hline $0 $   & $0 $  &$1 $   & $0 $   &  $ 0$   \\ \hline \hline
$b_{-4} b_{-3} c_{-2}c_1\ket{0} $   & $b_{-5} b_{-2} c_{-2}c_1\ket{0} $ 
 &$-b_{-6} c_{-2}\ket{0} $   & $b_{-5} b_{-3} c_{-1}c_1\ket{0} $   &  
$b_{-6} b_{-2} c_{-1}c_0\ket{0} $ 
\\ \hline $0 $   & $ -1$  &$0 $   & $0 $   &  $ 0$   \\ \hline \hline
$-b_{-7} c_{-1}\ket{0} $   & $b_{-3} b_{-2}  c_{-2} c_{-1}\ket{0} $  
&$b_{-5} b_{-4} c_{0} c_1\ket{0} $   & $b_{-6} b_{-3} c_{0}c_1\ket{0} $   
&  $b_{-7} b_{-2} c_{0}c_1\ket{0} $ 
\\ \hline $ -1$   & $-1 $  &$2 $   & $-2 $   &  $2 $   \\ \hline \hline
$-b_{-8}  c_{0}\ket{0} $   & $b_{-3} b_{-2} c_{-3} c_{0}\ket{0} $ 
 &$b_{-4} b_{-2} c_{-2} c_{0}\ket{0} $   & $b_{-4} b_{-3}  c_{-1} c_{0}\ket{0} $  
 &  $b_{-5} b_{-2} c_{-1} c_{0}\ket{0} $ 
\\ \hline $-2 $   & $2 $  &$-2 $   & $2 $   &  $-2 $   \\ \hline \hline
$L^m_{-8}\ket{0} $   & $b_{-2}c_1 L^m_{-7}\ket{0} $  &$b_{-3}c_1 L^m_{-6}\ket{0} $   
& $-b_{-2}  c_{0} L^m_{-6}\ket{0} $   &  $b_{-4}c_1 L^m_{-5}\ket{0} $ 
\\ \hline $-1/4 $   & $0 $  &$0 $   & $0 $   &  $0 $   \\ \hline \hline
$-b_{-2} c_{-1} L^m_{-5}\ket{0} $   & $-b_{-3}  c_{0} L^m_{-5}\ket{0} $  
&$b_{-5}c_1 L^m_{-4}\ket{0} $   & $-b_{-2} c_{-2} L^m_{-4}\ket{0} $   &  
$-b_{-3} c_{-1} L^m_{-4}\ket{0} $ 
\\ \hline $ 0$   & $ 0$  &$-1/2 $   & $-1/2 $   &  $ 1/2$   \\ \hline \hline
$b_{-3} b_{-2} c_{0}c_1 L^m_{-4}\ket{0} $   & $-b_{-4} c_{0} L^m_{-4}\ket{0} $  
&$L^m_{-4}L^m_{-4}\ket{0} $   & $b_{-6} c_1 L^m_{-3}\ket{0} $   &  
$-b_{-2} c_{-3} L^m_{-3}\ket{0} $ 
\\ \hline $ -1$   & $1 $  &$1/8 $   & $ 0$   &  $0 $   \\ \hline \hline
$-b_{-3}  c_{-2} L^m_{-3}\ket{0} $   & $b_{-3} b_{-2} c_{-1} c_1 L^m_{-3}\ket{0}$ 
 &$-b_{-4}  c_{-1} L^m_{-3}\ket{0} $   & $b_{-4} b_{-2} c_{0} c_1 L^m_{-3}\ket{0} $ 
  &  $-b_{-5}  c_{0} L^m_{-3}\ket{0} $ 
\\ \hline $ 0$   & $0 $  &$ 0$   & $ 0$   &  $ 0$   \\ \hline \hline
$L^m_{-5} L^m_{-3}\ket{0} $   & $b_{-2} c_1  L^m_{-4} L^m_{-3}\ket{0} $  &$b_{-3} c_1 L^m_{-3} L^m_{-3}\ket{0}
 $   & $-b_{-2} c_{0} (L^m_{-3})^2\ket{0} $   &  $b_{-7} c_1 L^m_{-2}\ket{0} $ 
\\ \hline $ 0$   & $0 $  &$0 $   & $ 0$   &  $-1 $   \\ \hline \hline
$-b_{-2}  c_{-4} L^m_{-2}\ket{0} $   & $-b_{-3}  c_{-3} L^m_{-2}\ket{0} $  
&$b_{-3} b_{-2} c_{-2} c_1 L^m_{-2}\ket{0} $   & $-b_{-4}  c_{-2} L^m_{-2}\ket{0} $ 
  &  $b_{-4} b_{-2} c_{-1} c_1 L^m_{-2}\ket{0} $ 
\\ \hline $ 0$   & $ -1$  &$ 1$   & $ 0$   &  $0 $   \\ \hline \hline
$-b_{-5}  c_{-1} L^m_{-2}\ket{0} $   & $b_{-4} b_{-3} c_{0} c_1 L^m_{-2}\ket{0} $  
&$b_{-5} b_{-2} c_{0} c_1 L^m_{-2}\ket{0} $   & $-b_{-6} c_{0} L^m_{-2}\ket{0} $  
 &  $b_{-3} b_{-2}  c_{-1} c_{0} L^m_{-2}\ket{0} $ 
\\ \hline $ 1$   & $2 $  &$ -2$   & $2 $   &  $ 2$   \\ \hline \hline
$L^m_{-6} L^m_{-2}\ket{0} $   & $b_{-2} c_1 L^m_{-5} L^m_{-2}\ket{0} $ 
 &$b_{-3} c_1 L^m_{-4} L^m_{-2}\ket{0} $   & $-b_{-2} c_{0} L^m_{-4} L^m_{-2}\ket{0} $ 
  &  $b_{-4} c_1  L^m_{-3} L^m_{-2}\ket{0} $ 
\\ \hline $ 0$   & $ 0$  &$ 1/2$   & $ -1$   &  $0 $   \\ \hline \hline
$-b_{-2}  c_{-1} L^m_{-3} L^m_{-2}\ket{0} $   & $-b_{-3}  c_{0} L^m_{-3} L^m_{-2}\ket{0} $ 
 &$(L^m_{-3})^2 L^m_{-2}\ket{0} $   & $b_{-5} c_1 (L^m_{-2})^2\ket{0} $  
 &  $-b_{-2} c_{-2} (L^m_{-2})^2 \ket{0} $ 
\\ \hline $ 0$   & $ 0$  &$0 $   & $1/2 $   &  $1/2 $   \\ \hline \hline
$ -b_{-3}  c_{-1} (L^m_{-2})^2 \ket{0}$   & $b_{-3} b_{-2} c_{0} c_1 (L^m_{-2})^2 \ket{0} $ 
 &$-b_{-4}  c_{0} (L^m_{-2})^2 \ket{0} $   & $L^m_{-4} (L^m_{-2})^2 \ket{0} $   &  
$b_{-2} c_1  L^m_{-3} (L^m_{-2})^2 \ket{0} $ 
\\ \hline $ -1/2$   & $1 $  &$ -1$   & $-1/4 $   &  $0 $   \\ \hline \hline
$b_{-3} c_1(L^m_{-2})^3\ket{0} $   & $-b_{-2} c_{0} (L^m_{-2})^3 \ket{0} $  
&$(L^m_{-2})^4 \ket{0} $   & $ $   &  $ $ 
\\ \hline $ -1/6$   & $1/3 $  &$ 1/24$   & $ $   &  $ $   \\ \hline \hline
\end{tabular} 
}
\ee
\section{The Proof for the Simplified Expression for the Identity}
In this section we wish to present the proof for the analytic
expression for the identity as given in (\ref{IdId}).
We remind the reader of the expression:
\begin{eqnarray}
\ket{\cal{I}} &=& 
  \left( \prod\limits_{n=2}^{\infty} 
	\exp\left\{- \frac{2}{2^n} L_{-2^n}\right\} \right)
  e^{L_{-2}} \ket{0} \nonumber \\
&=& 
\ldots
\exp(-\frac{2}{2^3} L_{-2^3}) \exp(-\frac{2}{2^2} L_{-2^2})
  \exp(L_{-2}) \ket{0},
\end{eqnarray}
or its BPZ conjugate form\footnote{Please notice that, besides the
	replacement $L_{n}\rightarrow (-)^n L_{-n}$, the orders under
	BPZ-conjugation are also reversed. This is because we use $L_n$
	instead of the oscillators $\alpha_m$, whose
	orders do not get reversed under BPZ.}
\begin{equation}
\bra{{\cal{I}}} = \bra{0}U_h U_{f_2}U_{f_3} U_{f_4} \ldots ,
\label{E:Iprod}
\end{equation}
where $U_{f_n}=e^{- \frac{2}{2^n} L_{2^n}}$ for $n\geq 2$ and 
$U_h=e^{L_{2}}$ . In \cite{0006240}, the identity is given by
$\bra{{\cal{I}}} =\bra{0} U_{f_{\cal I}}$ where $U_{f_{\cal I}}$ is
the operator corresponding to the function  
\[
  f_{\cal I} (z) = \frac{z}{1-z^2}.
\]
Using the composition law $U_{g_1} U_{g_2} = U_{g_1 \circ g_2}$, what
we  need is to prove  
\[
U_h U_{f_2}U_{f_3} U_{f_4} \ldots=U_{h \circ f_2 \circ f_3 \circ \ldots}
=U_{f_{\cal I}}
\]
which is equivalent to proving
\begin{equation}
\label{E:maineq}
  \lim_{k\to \infty} h \circ f_2 \circ \dots \circ f_k (z)= 
     f_{\cal I}(z)=\frac{z}{1-z^2}.
\end{equation}

For the operator $U_f=e^{aL_n}$, the corresponding function $f$ is given
by \cite{LPP}
\[
 f(z) = \exp \left\{ a z^{n+1} \partial_z \right\} z = 
   \frac{z}{(1-a n z^n)^{1/n}}~,
\]
so we have
\[
\begin{array}{rcl}
 h(z) &=&\frac{z}{(1-2 z^2)^{1/2}}\\
 f_n(z) &=& \frac{z}{(1+ 2 z^{2^n})^{1/2^n}}.
\end{array}
\]
A useful property of the $f_n$ is that $f_n(z) = (g(z^{2^n}))^{1/2^n}$ where
\[
 g(z) := \frac{z}{1+ 2 z} = \frac{1}{2 + 1/z}.
\]

Before writing down the general form, first let us do an example:
\begin{eqnarray*}
f_2\circ f_3 \circ f_4(z) & = & f_2\circ f_3[ (g[z^{2^4}])^{1/2^4}]
\\
& = & f_2[( g[((g[z^{2^4}])^{1/2^4})^{2^3}])^{1/2^3}] \\
& = & f_2[( g[ g^{1/2}[z^{2^4}]])^{1/2^3}] \\
& = & (g[ (( g[ g^{1/2}[z^{2^4}]])^{1/2^3})^{2^2}])^{1/2^2} \\
& = & (g[g^{1/2}[ g^{1/2}[z^{2^4}]]])^{1/2^2} \\
& = & (g^{1/2}[g^{1/2}[ g^{1/2}[z^{2^4}]]])^{1/2}.
\end{eqnarray*}
Now it is easy to see that the general form is 
\[
 h \circ f_{2} \circ f_3  \circ \ldots \circ f_{k+1} (z) = 
  h\circ (
   \underbrace{g^{\frac{1}{2}} \circ \ldots \circ 
     g^{\frac{1}{2}}}_{k}(z^{2^{k+1}}))^{\frac12}.
\]
Thus equation (\ref{E:maineq}) is equivalent to showing
that
\[\lim_{k\to\infty}
 \underbrace{g^{\frac{1}{2}} \circ \ldots \circ 
  g^{\frac{1}{2}}}_{k}(z^{2^{k+1}}) = (h^{-1}(f(z)))^2 = \frac{z^2}{1+z^4}.
\]
The left hand side can be written as
\[
 (\underbrace{(2+(2+ \dots + (2}_{k}+ 1/z^{2^{k+1}}
 )^{\frac{1}{2}}\ldots 
 )^{\frac{1}{2}}
 )^{\frac{1}{2}})^{-1} =
 z^2 ((2 z^{2^2} + (2 z^{2^3} + \ldots (2 z^{2^{k +1}} +1)^{\frac{1}{2}}
 \ldots )^{\frac{1}{2}})^{\frac{1}{2}})^{-1}.
\]
Thus (\ref{E:maineq}) reduces to the verification of the equation
\[
 \lim_{k\to \infty}
(2 z^{2^2} + (2 z^{2^3} + \ldots (2 z^{2^{k +1}} +1)^{\frac{1}{2}}
 \ldots )^{\frac{1}{2}})^{\frac{1}{2}} = 1+z^{2^2}.
\]
This can be done as follows.  Consider first squaring both sides of the 
above equation and canceling $2 z^{2^2}$ from the two sides, we get 
\[
 \lim_{k\to \infty}
 (2 z^{2^3} + \ldots (2 z^{2^{k +1}} +1)^{\frac{1}{2}}
 \ldots )^{\frac{1}{2}})^{\frac{1}{2}} = 1+z^{2^3}.
\]
Repeating the above operation $k$ times,  the left hand side gives $1$
while the right hand side gives $1+z^{2^{k+2}}$.  
Thus as long
as $z < 1$, we get that the left and right hand sides do converge to 
each other as $k \to \infty$.

\bibliographystyle{JHEP}

\end{document}